\documentclass[a4paper]{article}
\usepackage[14pt]{extsizes}
\usepackage[english]{babel}
\usepackage[intlimits]{amsmath} 
\usepackage{hyperref}
\usepackage{indentfirst}
\usepackage{geometry} 
\usepackage{xfrac}
\usepackage{cite}
\usepackage{setspace}
\providecommand{\keywords}[1]{\textbf{Keywords: } #1}

\begin{document}
\title{An improvement of D-MORPH method for finding quantum optimal control\thanks{Punlished in "Mezdunarodnyj naucno-issledovatel'skij zurnal" ["International Research Journal"], 2016, Vol.~6(48), p. 94--99. DOI: 10.18454/IRJ.2016.48.057 [in Russian]}}
\author{Konstantin Zhdanov\thanks{Department of applied mathematics and control processes, Saint-Petersburg state university, Russia. ORCID: 0000-0003-2290-6324}}
\date{June, 2016}

\maketitle
\begin{abstract}
The paper examines the prominent algorithm D-MORPH to search for the optimal control of a quantum system in order to implement desired unitary evolution of the quantum system at the final time, and reveals new mathematical expressions for various orders’ corrections to the algorithm, that include information about the commutators of the system’s Hamiltonian. Inclusion of such corrections results in faster optimal quantum control’s search with high precision, i.e. allows saving of computational resources.
\end{abstract}


\section{Introduction}
Many papers of distinguished specialists in the area of quantum control~\cite{Moore1, Moore2, Moore3, Riviello1, Riviello2} use the D-MORPH method to construct optimal quantum control. The method is based on introducing a new variable \(s\) which represents the optimization algorithm's progress towards the minimum of an objective function. Variable \(s\) has an interesting property: while increasing \(s\), the optimization algorithm is constructing new controls that drive the system closer and closer to the desired state associated with the minimum of the objective function.

This optimization algorithm is approximate since it's based on solving a system of ordinary differential equations numerically with the help of the popular MATLAB's ode45 method, i.~e. a variable-step Runge--Kutta's method of order four. As it's been demonstrated many times~\cite{Moore1, Moore2, Moore3, Riviello1, Riviello2}, this optimization algorithm allows construction of controls resulting in close-to-the-optimal values of the objective function. One weakness of the algorithm is the requirement to solve the system on a large interval of integration \([0,S]\) in order to achieve a high accuracy, which can be very computationally demanding.

This paper introduces a way to improve the aforementioned algorithm --- new formulas for corrections of different orders to the original method are derived so that they allow high-accuracy reaching of the minimum of an objective function, which represents the distance between the actual evolution of a quantum system and the desired evolution, with fewer integration steps than the original method requires and so spending less time. These corrections allow searching for optimal quantum controls to be effectively carried out on common computers with moderate performance.
\section{The original method and corrections to it}
Let's consider an \(N\)-level quantum system subjected to \(n\) control fields and formulate a task: find such control fields \(\epsilon_k(t)\), \(k=\overline{1,n}\), that the quantum system at a moment \(T\) implements the desired evolution operator \(U_D\) as accurately as possible. The accuracy is expressed in terms of an objective function \(J\), which represents the distance between the actual evolution \(U(T,0)\) of the system and the desired evolution \(U_D\).

The original algorithm is formulated as follows. A new variable \(s\) is introduced such that objective \(J(s)\) and controls \(\epsilon_k(s, t)\) both depend on it. Then we find the derivative
\[
\frac{\mathrm{d} J}{\mathrm{d} s} = \sum_{k=1}^n \int_0^T \frac{\partial J(s)}{\partial \epsilon_k(s,t)} \frac{\mathrm{d}\epsilon_k(s,t)}{\mathrm{d} s} \mathrm{d} t.
\]
To find the minimum of the objective function \(J\), it's sufficient to require that \(\sfrac{\mathrm{d}J}{\mathrm{d}s} \leq 0\) holds. In that case, if we increase \(s\), \(J\) at least won't increase.
\medskip

Let's divide the interval \([0,T]\) into \(L\) subintervals of the same length \(\Delta t\) and treat controls as piecewise constant with respect to \(t\) functions
\[
\epsilon_k(s,t) = \sum_{l=1}^L \epsilon_k^l(s) \chi_{[t_{l-1}, t_l]}(t),
\]
where \(\chi_{[t_{l-1}, t_l]}(t)\) --- the indicator function of interval \([t_{l-1}, t_l]\), 
\[
\chi_{[t_{l-1}, t_l]}(t) = 
    \begin{cases}
      1, & t\in [t_{l-1}, t_l] \\
      0, & t \notin [t_{l-1}, t_l]
    \end{cases}.
\]
Then inequality \(\sfrac{\mathrm{d}J}{\mathrm{d}s} \leq 0\) can be written as
\[
\frac{\mathrm{d} J}{\mathrm{d} s} = \sum_{k=1}^n \sum_{l=1}^L \int_{t_{l-1}}^{t_l} \frac{\partial J(s)}{\partial \epsilon_k(s,t)} \frac{\mathrm{d}\epsilon_k^l(s)}{\mathrm{d} s} \mathrm{d} t \leq 0.
\]
The inequality is satisfied if we let
\begin{equation}\label{DMORPH_cond}
\frac{\mathrm{d}\epsilon_k^l(s)}{\mathrm{d} s} = -\frac{\partial J(s)}{\partial \epsilon_k(s,t)}, \quad t\in [t_{l-1}, t_l], \quad l=\overline{1,L},\quad k=\overline{1,n}.
\end{equation}

Therefore the original method solves the system of differential equations~\eqref{DMORPH_cond}, where \(\epsilon_k^l(s)\) are unknown functions to be found. As it's assumed that a functional form of \(J\) with respect to \(\epsilon_k(s, t)\) is known in advance, the right-hand side of the expression above is assumed to be known as well. It's worth noting that the right-hand side depends on all \(\epsilon_k^l(s)\).
	  
In papers~\cite{Moore1, Moore2, Moore3, Riviello1, Riviello2} such systems of differential equations are solved with a variable-step Runge--Kutta's method of order four (MATLAB's ode45) on \([0,S]\), where \(S\) --- a large number. Then an approximately optimal control is retrieved at \(s=S\) owing to the method guarantees that the objective function at least doesn't increase while increasing \(s\). In this paper the aforementioned original method is modified so as to get approximate optimal control with given accuracy by solving the system on smaller interval \([0,S]\) than used in the original method. A decrease in the length of the interval of integration should diminish the time spent by ode45 method, which should be beneficial when using computers with moderate performance.

Since the right-hand side of system~\eqref{DMORPH_cond} depends on time while the left-hand side doesn't, we integrate each equation of the system with respect to \(t\) from \(t_{l-1}\) to \(t_l\).
\[
\int_{t_{l-1}}^{t_l}\frac{\mathrm{d}\epsilon_k^l(s)}{\mathrm{d} s} \mathrm{d} t = -\int_{t_{l-1}}^{t_l}\frac{\partial J(s)}{\partial \epsilon_k(s,t)} \mathrm{d} t, \quad l=\overline{1,L},\quad k=\overline{1,n},
\]
\[
\frac{\mathrm{d}\epsilon_k^l(s)}{\mathrm{d} s} = -\frac{1}{\Delta t}\int_{t_{l-1}}^{t_l}\frac{\partial J(s)}{\partial \epsilon_k(s,t)} \mathrm{d} t, \quad l=\overline{1,L},\quad k=\overline{1,n}.
\]

In what follows, we are going to expand the integral on the right-hand side of the expression above. For a task of implementing the desired evolution operator, the objective function in the original method is taken as~\cite{Moore1, Moore3}
\[
J = \frac{1}{2} - \frac{1}{2N}\operatorname{Re}\operatorname{Tr}\left[U_D^* U(T,0) \right].
\]

This objective function represents the distance between the desired evolution operator \(U_D\) and the actual evolution operator \(U(T,0)\) at time \(T\). The derivative of \(J\) with respect to controls is written as~\cite{Moore3}
\[
\frac{\partial J(s)}{\partial \epsilon_k(s,t)} = -\frac{1}{2N} \operatorname{Im}\operatorname{Tr}\left[U_D^* U(T,t)H_k U(t,0) \right],
\]
while the integral of this expression is equal to
\begin{multline*}
\int_{t_{l-1}}^{t_l}\frac{\partial J(s)}{\partial \epsilon_k(s,t)} \mathrm{d} t = -\frac{1}{2N} \operatorname{Im}\operatorname{Tr}\left[U_D^* \int_{t_{l-1}}^{t_l}U(T,t)H_k U(t,0)\mathrm{d} t \right] =\\
= -\frac{1}{2N} \operatorname{Im}\operatorname{Tr}\left[U_D^* U(T,0)\int_{t_{l-1}}^{t_l}U^*(t,0)H_k U(t,0)\mathrm{d} t \right].
\end{multline*}

In general, the evolution operator of a quantum system with Hamiltonian \(H(s,t)=H_0 + \sum_{k=1}^n \epsilon_k(s,t)H_k\) is expressed as (\(\hbar = 1\))~\cite{Moore1}

\[
U(T,0) = \operatorname{T}\exp\left\{-\imath \int_{0}^{T} H(s, \tau) \mathrm{d} \tau\right\},
\]
where \(\operatorname{T}\) --- time-ordering operator. Since the controls are piecewise constant, the evolution operator turns into a composition of matrix exponentials
\begin{multline*}
U(T,0) = \exp\left\{-\imath \int_{t_{L-1}}^{T} H(s, \tau) \mathrm{d} \tau\right\} \exp\left\{-\imath \int_{t_{L-2}}^{t_{L-1}} H(s, \tau) \mathrm{d} \tau\right\} \cdots \\
\cdots\exp\left\{-\imath \int_{0}^{t_1} H(s, \tau) \mathrm{d} \tau\right\}.
\end{multline*}
Using the following properties of the evolution operator
\[
U(t,0) = U(t,t_{l-1}) \cdots U(t_{1},0) = U(t,t_{l-1})U(t_{l-1},0),\quad t \in [t_{l-1}, t_l],
\]
\[
U^*(t,0) = U^*(t_{1},0)\cdots U^*(t,t_{l-1})  = U^*(t_{l-1},0)U^*(t,t_{l-1}),\quad t \in [t_{l-1}, t_l]
\]
we get
\[
\int_{t_{l-1}}^{t_l}U^*(t,0)H_k U(t,0)\mathrm{d} t = U^*(t_{l-1}, 0) \int_{t_{l-1}}^{t_l}U^*(t,t_{l-1})H_k U(t,t_{l-1})\mathrm{d} t U(t_{l-1}, 0).
\]
Let's express the evolution operator as the matrix exponential.
\begin{multline*}
\int_{t_{l-1}}^{t_l}U^*(t,t_{l-1})H_k U(t,t_{l-1})\mathrm{d} t =\\
= \int_{t_{l-1}}^{t_l} \exp\left[(t-t_{l-1}) X\right] H_k \exp\left[-(t-t_{l-1}) X\right] \mathrm{d} t
=\int_{t_{l-1}}^{t_l} \operatorname{Ad}_{\exp[(t-t_{l-1})X]}H_k\mathrm{d} t,
\end{multline*}
where \(X=\imath\left(H_0 + \sum_{k=1}^n \epsilon_k^l H_k \right)\), \(\operatorname{Ad}_{A}B = A B A^{-1}\). It's well known in the general theory of Lie algebras that
\[
\operatorname{Ad}_{\exp(A)}B = \exp(\operatorname{ad}_A) \circ B,
\]
where \(\operatorname{ad}_A = A B - B A\). Using this expression we get
\[
\int_{t_{l-1}}^{t_l} \operatorname{Ad}_{\exp[(t-t_{l-1})X]}H_k\mathrm{d} t = \int_{t_{l-1}}^{t_l} \exp\left(\operatorname{ad}_{(t-t_{l-1})X}\right)\circ H_k\mathrm{d} t
\]
Let's expand operator \(\exp\left(\operatorname{ad}_{(t-t_{l-1})X}\right)\) in a Taylor series and use identity \(\operatorname{ad}_{(t-t_{l-1})X} = (t-t_{l-1}) \operatorname{ad}_{X}\)
\[
\exp\left(\operatorname{ad}_{(t-t_{l-1})X}\right) = \sum_{j=0}^{\infty} \frac{(t-t_{l-1})^j}{j!} \operatorname{ad}_{X}^j.
\]
Then we integrate it
\[
\int_{t_{l-1}}^{t_l} \exp\left(\operatorname{ad}_{(t-t_{l-1})X}\right)\circ H_k\mathrm{d} t = \Delta t \left(H_k + \frac{\Delta t}{2} [X, H_k] + \frac{(\Delta t)^2}{3!} [X, [X, H_k]] + \cdots \right).
\]
Finally we get
\begin{multline*}
\int_{t_{l-1}}^{t_l}\frac{\partial J(s)}{\partial \epsilon_k(s,t)} \mathrm{d} t = -\frac{\Delta t}{2N} \operatorname{Im}\operatorname{Tr}\left[U_D^* U(T,t_{l-1})\left(H_k + \frac{\Delta t}{2} [\imath H(t_l), H_k] + \right.\right.\\
\left.\left.+ \frac{(\Delta t)^2}{3!} [\imath H(t_l), [\imath H(t_l), H_k]] + \cdots \right) U(t_{l-1},0) \right].
\end{multline*}
 
Then, system~\eqref{DMORPH_cond} can be written in its final form as
\begin{multline*}
\frac{\mathrm{d} \epsilon_k^l}{\mathrm{d} s} = \frac{1}{2N} \operatorname{Im}\operatorname{Tr}\left[U_D^* U(T,t_{l-1})\left(H_k + \frac{\Delta t}{2} [\imath H(t_l), H_k] + \right.\right.\\
\left.\left.+ \frac{(\Delta t)^2}{3!} [\imath H(t_l), [\imath H(t_l), H_k]] + \cdots \right) U(t_{l-1},0) \right], \quad k=\overline{1,n},\quad l=\overline{1,L}.
\end{multline*}

Numerically solving this system on interval \([0,S]\) with the MATLAB's ode45 should give controls that implement the desired evolution operator more accurately than the original method. This improvement could be accounted for by the use of information about the Hamiltonian's commutators.
\bigskip

It's expected that the more terms of the series on the right-hand side of the expression above we use during computations, the more accurate we get approximately optimal control. Of course, when using terms of order \((\Delta t)^j\), we have to compute repeated commutators of order \(j\) and this can be very computationally demanding in general. For this reason, a maximum order of terms to use on the right-hand side should be chosen such that the speed-up of the new algorithm beats a slow-down induced by the commutators' computation.
\section{Numerical experiment}
To confirm the expected speed-up provided by the new formula, a quantum system composed of two spin-\(\sfrac{1}{2}\) particles and described by the following dimensionless Hamiltonian (\(\hbar = 1\)) was taken.
\[
H = \sum_{i=1}^2 S_z^i \omega_i + \sum_{k=1}^2 \epsilon_k S_x^k + C_x^{(12)} S_x^1 S_x^2 + C_y^{(12)} S_y^1 S_y^2 + C_z^{(12)} S_z^1 S_z^2,
\]
where \(\omega_1=20\), \(\omega_2=30\), \(C_x^{(12)}=110\), \(C_y^{(12)}=120\), \(C_z^{(12)}=130\), \(S_i^1 = S_i \otimes I\), \(S_i^2 = I\otimes S_i\), \(i=x,y,z\),
\[
S_x = \frac{1}{\sqrt{2}}\begin{pmatrix}
0 & 1\\
1 & 0
\end{pmatrix},
S_y = \frac{1}{\sqrt{2}}\begin{pmatrix}
0 & -\imath \\
\imath & 0
\end{pmatrix},
S_z = \frac{1}{\sqrt{2}}\begin{pmatrix}
1 & 0\\
0 & -1
\end{pmatrix},
I = \begin{pmatrix}
1 & 0\\
0 & 1
\end{pmatrix}.
\]
The task was to implement the following operators as accurate as possible
\[
U_D^1 = \mathrm{e}^{\frac{\imath \pi}{4}}\begin{pmatrix}
1 & 0 & 0 & 0\\
0 & 1 & 0 & 0\\
0 & 0 & 0 & 1\\
0 & 0 & 1 & 0
\end{pmatrix},\quad
U_D^2 = \mathrm{e}^{\frac{\imath \pi}{4}}\begin{pmatrix}
1 & 0 & 0 & 0\\
0 & 0 & 1 & 0\\
0 & 1 & 0 & 0\\
0 & 0 & 0 & 1
\end{pmatrix},
\]
which are respectively the CNOT quantum gate (controlled NOT) and the SWAP quantum gate (a swap of qubits' values).

Two models were compared to each other:
\begin{equation} \label{DMORPH}
\frac{\mathrm{d}\epsilon_k^l}{\mathrm{d} s} = \frac{1}{2N}\operatorname*{Im} \operatorname*{Tr}\left(U_D^* U(T, t_{l-1})H_k U(t_{l-1}, 0)\right),
\end{equation}
\begin{equation} \label{Corrections_1st}
\frac{\mathrm{d}\epsilon_k^l}{\mathrm{d} s} = \frac{1}{2N}\operatorname*{Im} \operatorname*{Tr}\left(U_D^* U(T, t_{l-1})  \left\{H_k + \frac{\imath \Delta t}{2} \left[H(t_l),H_k\right] \right\} U(t_{l-1}, 0)\right).
\end{equation}

To numerically solve these systems, the MATLAB's ode45 method was employed with the absolute error tolerance set to \(10^{-4}\). For the CNOT gate, the initial controls were set to zero, whereas for the SWAP gate the initial controls were taken by sampling the function \(10^{-5}\sin(\sfrac{t}{T})\) owing to the fact that zero initial controls didn't allow the optimization to converge.

The special structure of the given Hamiltonian allows a simplification:
\[
[H(t), H_k] = [H_0 + \sum_{i=1}^2 \epsilon_i(t) H_i, H_k] = [H_0, H_k].
\]

The final time \(T\) was set to several values: \(0.5\), \(1\), \(5\), \(10\).
The number of subintervals \(L\) was taken as \(150\), \(300\). The length of the interval of integration \(S\) was taken as the multiples of \(100\):  \(100\), \(200\), \(300\), \(400\), ...
The times required to numerically solve the aforementioned models on \([0,S]\) with the error in the gates' implementation not greater than \(10^{-7}\) were compared to each other. Computations were performed on a laptop with a quad-core Intel Core i7-4702MQ 2.20 GHz processor and 8 Gb RAM. Results of the comparison for the CNOT gate are in tables 1 and 2, for the SWAP gate --- in tables 3 and 4. The time spent to solve the systems was measured with the MATLAB's functions tic() and toc().

Table 1 --- Results for the CNOT gate.
\begin{center}
\begin{tabular}{ *{9}{|c} |}
\hline
& \multicolumn{2}{|c|}{\(T=10\)} & \multicolumn{2}{|c|}{\(T=10\)} & \multicolumn{2}{|c|}{\(T=5\)} & \multicolumn{2}{|c|}{\(T=5\)}\\
& \multicolumn{2}{|c|}{\(L=300\)} & \multicolumn{2}{|c|}{\(L=150\)} & \multicolumn{2}{|c|}{\(L=300\)} & \multicolumn{2}{|c|}{\(L=150\)}\\
\hline
Method & \eqref{DMORPH} & \eqref{Corrections_1st} & \eqref{DMORPH} & \eqref{Corrections_1st} & \eqref{DMORPH} & \eqref{Corrections_1st} & \eqref{DMORPH} & \eqref{Corrections_1st}\\
\hline
S & 400 & 100 & x* & 600 & 900 & 200 & 1200 & 400\\
\hline
Time, sec. & 73.76 & 35.76 & x & 46.98 & 97.53 & 39.17 & 29.92 & 20.27\\
\hline
Error (\(\times 10^{-8}\)) & \(0.409\) & \(3.94\) & x & \(0.412\) & \(1.007\) & \(1.85\) & \(3.54\) & \(3.19\)\\
\hline
\end{tabular}
\end{center}
*Method didn't converge.

Table 2 --- Results for the CNOT gate (contrinued)
\begin{center}
\begin{tabular}{ *{9}{|c} |}
\hline
& \multicolumn{2}{|c|}{\(T=1\)} & \multicolumn{2}{|c|}{\(T=1\)} & \multicolumn{2}{|c|}{\(T=0.5\)} & \multicolumn{2}{|c|}{\(T=0.5\)}\\
& \multicolumn{2}{|c|}{\(L=300\)} & \multicolumn{2}{|c|}{\(L=150\)} & \multicolumn{2}{|c|}{\(L=300\)} & \multicolumn{2}{|c|}{\(L=150\)}\\
\hline
Method & \eqref{DMORPH} & \eqref{Corrections_1st} & \eqref{DMORPH} & \eqref{Corrections_1st} & \eqref{DMORPH} & \eqref{Corrections_1st} & \eqref{DMORPH} & \eqref{Corrections_1st}\\
\hline
S & 1000 & 800 & 1000 & 700 & 3900 & 3600 & 4000 & 3600\\
\hline
Time, sec. & 36.68 & 37.66 & 11.66 & 10.39 & 66.22 & 63.21 & 19.52 & 18.6\\
\hline
Error (\(\times 10^{-8}\)) & \(2.2\) & \(1.4\) & \(3.36\) & \(5.1\) & \(9.36\) & \(9.41\) & \(7.80\) & \(6.93\)\\
\hline
\end{tabular}
\end{center}

Table 3 --- Results for the SWAP gate.
\begin{center}
\begin{tabular}{ *{9}{|c} |}
\hline
& \multicolumn{2}{|c|}{\(T=10\)} & \multicolumn{2}{|c|}{\(T=10\)} & \multicolumn{2}{|c|}{\(T=5\)} & \multicolumn{2}{|c|}{\(T=5\)}\\
& \multicolumn{2}{|c|}{\(L=300\)} & \multicolumn{2}{|c|}{\(L=150\)} & \multicolumn{2}{|c|}{\(L=300\)} & \multicolumn{2}{|c|}{\(L=150\)}\\
\hline
Method & \eqref{DMORPH} & \eqref{Corrections_1st} & \eqref{DMORPH} & \eqref{Corrections_1st} & \eqref{DMORPH} & \eqref{Corrections_1st} & \eqref{DMORPH} & \eqref{Corrections_1st}\\
\hline
S & 900	& 300 &	x* & 300 & 3200 & 400 & 1900 & 800\\
\hline
Time, sec. & 143.96 & 85.99 & x & 21.52 & 245.22 & 46.68 & 35.13 & 19.31\\
\hline
Error (\(\times 10^{-8}\)) & \(1.08\) & \(0.0001\) & x & \(0.0026\) & \(4.79\) & \(8.14\) & \(8.43\) & \(0.237\)\\
\hline
\end{tabular}
\end{center} 
*Method didn't converge.

Table 4 --- Results for the SWAP gate (continued)
\begin{center}
\begin{tabular}{ *{9}{|c} |}
\hline
& \multicolumn{2}{|c|}{\(T=1\)} & \multicolumn{2}{|c|}{\(T=1\)} & \multicolumn{2}{|c|}{\(T=0.5\)} & \multicolumn{2}{|c|}{\(T=0.5\)}\\
& \multicolumn{2}{|c|}{\(L=300\)} & \multicolumn{2}{|c|}{\(L=150\)} & \multicolumn{2}{|c|}{\(L=300\)} & \multicolumn{2}{|c|}{\(L=150\)}\\
\hline
Method & \eqref{DMORPH} & \eqref{Corrections_1st} & \eqref{DMORPH} & \eqref{Corrections_1st} & \eqref{DMORPH} & \eqref{Corrections_1st} & \eqref{DMORPH} & \eqref{Corrections_1st}\\
\hline
S & 2700 & 2600 & 2900 & 2500 & 3100 & 3200 & 3400 & 3200\\
\hline
Time, sec. & 67.25 & 65.67 & 22.22 & 22.77 & 52.05 & 48.67 & 17.69 & 16.01\\
\hline
Error (\(\times 10^{-8}\)) & \(9.005\) & \(7.57\) & \(9.57\) & \(6.9\) & \(9.42\) & \(4.24\) & \(9.42\) & \(4.94\)\\
\hline
\end{tabular}
\end{center}

For \(T=0.01\) and \(T=0.001\), the both methods didn't give accurate results, which perfectly agrees with the fact that in similar tasks there is a lower bound on time to construct control fields~\cite{Riviello2}.

These results show that the improved method proposed in this paper almost always achieves the desired accuracy in less time and on a smaller interval \([0, S]\) than the original method. This improvement is noticeable only for large \(T\) and large \(\Delta t\), which agrees with the corrections being proportional to \(\Delta t\) and hence they being small given small \(\Delta t\). Moreover, for some large \(\Delta t\) the original method doesn't converge at all whereas the improved method achieves the desired accuracy.
\section{Conclusion}
New formulas for constructing optimal controls in a quantum system were obtained in the form of corrections to the original D-MORPH method~\cite{Moore1, Moore2, Moore3, Riviello1, Riviello2}. By numerically solving a problem in the area of quantum computations, it was shown that the new formulas do speed up computations compared to the original method, i.~e. they allow implementation of the desired evolution operator with high accuracy with fewer ode45's steps and in less time, even if only one correcting term proportional to \(\Delta t\) is used. Thus, inclusion of additional information about the system's Hamiltonian (commutators) had a positive impact on accuracy achieved and on time spent, which is an advantage for finding optimal controls on computers with moderate performance.

\bibliographystyle{plain}

\end{document}